# The State-of-the-Art in AI-Based Malware Detection Techniques: A Review

Adam Wolsey

**Abstract**

Artificial Intelligence techniques have evolved rapidly in recent years, revolutionising the approaches used to fight against cybercriminals. But as the cyber security field has progressed, so has malware development, making it an economic imperative to strengthen businesses' defensive capability against malware attacks. This review aims to outline the state-of-the-art AI techniques used in malware detection and prevention, providing an in-depth analysis of the latest studies in this field. The algorithms investigated consist of Shallow Learning, Deep Learning and Bio-Inspired Computing, applied to a variety of platforms, such as PC, cloud, Android and IoT. This survey also touches on the rapid adoption of AI by cybercriminals as a means to create ever more advanced malware and exploit the AI algorithms designed to defend against them.

## 1. Introduction

The coronavirus pandemic has dramatically increased worldwide cyber-attacks, a phenomenon that has increasingly been termed the 'cyber pandemic' [1] and is expected to reach USD 10.5 trillion in annual damage costs by 2025 [2]. At the same time, the new business model of 'working from home' imposed by the pandemic has substantially increased most organisations' threat exposure [3].

More alarmingly, prior to the pandemic, an estimated 20% of cyber attackers used previously unseen malware or attack techniques, with many of these consisting of machine learning models that adapt to the environment to remain undetected. This proportion rose to 35% during the pandemic [4]. Combined, these trends indicate a negative direction in cybercrime, set to impact a large segment of global businesses in all industries and expected to advance at an unprecedented rate in the next decade.

With the exponential growth of technology, data and computing power, it is fundamental that more sophisticated tools are used to tackle rising modern problems. As humans cannot handle the growing complexity on their own, the dependency on AI has become unavoidable. It is predicted that the market for AI within cybersecurity will grow from USD 3.92 Billion in 2017 to USD 34.81 Billion by 2025 [5].

Furthermore, a survey conducted by the Capgemini Research Institute found that 69% of organisations think AI is necessary to respond to cyberattacks [6]. Now more than ever, AI is attracting greater attention from the public and private sectors. However, its power will inevitably fall into the hands of cybercriminals, creating the next generation of AI-powered malware.

This unprecedented, AI-powered upsurge in cybercriminal capabilities makes it all the more important for security experts to identify the types of detected malware swiftly and accurately. But despite significant advances in AI-driven malware detection methods, the current rate of progress is inadequate, and greater efforts to outpace cybercriminals are required.

Malware (short for malicious software) comes in many categories – such as viruses, worms, spyware, trojans, ransomware, etc. – but almost always has the goal of compromising systems and data or holding a victim to ransom.

Traditionally, malware detection was based entirely on comparing continuous byte sequences (called 'signatures') of a suspected malware file to the signatures of known malware held in a database. Over time, as newer, 'polymorphic' malware appeared, signature-based detection became less effective and was superseded by next-generation, heuristic-based and behavioural-based detection methods relying on machine learning models [7]. Currently, all the top malware-detection solutions (also called Endpoint Detection and Response – EDR) are underpinned by machine learning algorithms [8].

This literature review aims to present and analyse the latest efforts in developing novel, more effective ways to use artificial intelligence for malware detection, with the aim of providing comprehensive guidance for subsequent research.

This review can, therefore, assist researchers in developing an understanding of the malware detection field, as well as the new developments and research directions explored by the scientific community to address this complex challenge.

There are numerous papers outlining AI-based malware detection techniques. However, because this research focuses on the most recent trends in AI-based malware detection, papers older than 2016 will not be included in the scope of this survey.

The remainder of this paper is structured as follows: Sections 2 and 3 provide an overview of the main approaches to malware analysis and detection; Section 4 defines feature extraction and selection; Sections 5, 6 and 7 review the most recent papers using shallow learning, deep learning and bio-inspired AI algorithms for malware detection in host-based, cloud and IoT environments; Section 8 presents the state-of-the-art machine learning approaches for Android malware detection; and Section 9 describes malicious uses of AI for the purpose of designing malware. Finally, Section 10 provides conclusions, identifying potential challenges and future directions for the use of AI in malware detection.

## 2. Malware Analysis Approaches

Malware analysis is the foundation of malware detection and essential for developing effective malware detection techniques. Without malware analysis, which provides insight into the classification and functionality of the malicious file, detecting malware could not be achieved. Malware analysis is divided into static, dynamic and hybrid approaches, as described below [9]:

- **Static analysis**, whereby a suspected malicious file is inspected and analysed without executing it based on extracted low-level information such as system calls, the control flow graph, and the data flow graph. Static analysis produces a low number of false positives; however, it fails to detect unknown malware that uses code obfuscation.

- **Dynamic analysis**, whereby a suspected malicious file is inspected at runtime, usually within a sandbox (an isolated virtual machine designed for testing purposes, where malware can be executed without affecting system resources). The advantage is that the malware can be executed and analysed; therefore, unknown malware can be successfully detected. However, dynamic analysis is time-consuming and produces a high level of false positives.

- **Hybrid analysis**, whereby characteristics of static and dynamic analysis are combined to overcome the challenges of the two.

This paper will refer to these three types of analysis to describe the basis of the methods employed in state-of-the-art machine learning-based malware detection systems.

## 3. Malware Detection Techniques

As previously mentioned, malware detection techniques can be classified into three broad categories: signature-based, heuristic-based, and behaviour-based. These methods rely on results from malware analysis, and each method has its unique advantages and challenges [9]:

- **Signature-based detection** – this technique uses a known list of indicators of compromise (IOCs), which include specific byte sequences, API calls, file hashes, malicious domains or network attack patterns. Signature-based detection is, however, incapable of detecting previously unknown or encrypted malware and does not require machine learning models.

- **Behavioural-based detection** – involves monitoring a suspected executable file in an isolated environment and collecting all exhibited behaviours, then using methods of extracting useful features by which a machine learning model can classify the malicious behaviour.

- **Heuristic-based detection** – this technique relies on generating rules based on the results of the static/dynamic analysis to guide the inspection of the extracted data to support the proposed malware detection model. Such rules can either be generated manually (relying on the expertise of the security analysts) or automatically, using machine learning or tools such as YARA.

**4. Feature Extraction and Selection**

Throughout this paper, the term 'features' will be frequently used in relation to various machine learning algorithms. In machine learning, feature extraction or feature engineering is the process of transforming raw data into numerical features that can be 'understood' by the machine learning algorithm. Feature extraction is necessary to improve the model's effectiveness, as applying machine learning directly to raw data is generally ineffective.

On the other hand, feature selection is the process of removing unnecessary features to assist with developing a predictive model.

In a machine learning-based malware detection system, feature extraction and selection are critical steps in developing the system. Tables II, IV, V and VI below detail the features extracted by the authors of each paper evaluated in this review.

**5. Shallow Learning-Based Classification Methods for Malware Detection**

Shallow learning (SL) generally comprises the majority of machine learning models proposed prior to 2006 and, more specifically, any machine learning models not classified as deep learning. SL approaches traditionally depend heavily on features manually designed to solve a given task [10].

Nevertheless, shallow learning algorithms are still widely used, including in cyber security more broadly and malware detection in particular, as seen in Table II.

Fig. 1 below lists the most commonly used SL-based classification methods.

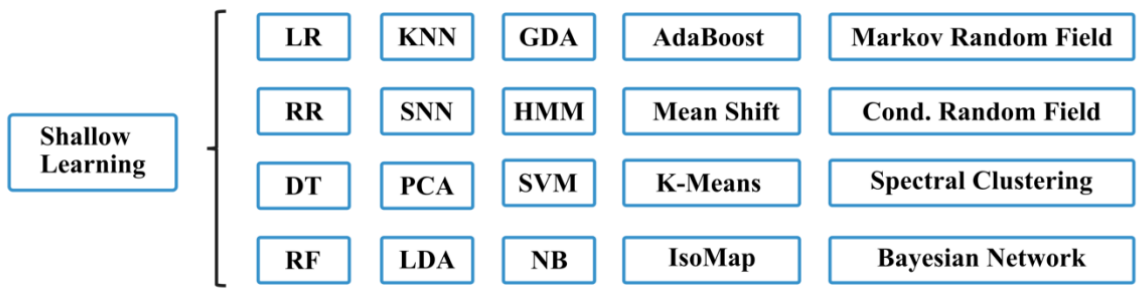

Fig. 1 Common shallow learning algorithms

Table I provides explanations for the acronyms used in Fig. 1:

Table I

Acronyms used in Fig. 1

| | | | |
|---|---|---|---|
| LR | Logistic Regression | PCA | Principal Component Analysis |
| RR | Ridge Regression | LDA | Linear Discriminant Analysis |
| DT | Decision Trees | GDA | Gaussian Discriminant Analysis |
| RF | Random Forest | HMM | Hidden Markov Models |
| KNN | K-Nearest Neighbour | SVM | Support-Vector Machines |
| SNN | Shallow Neural Network | NB | Naïve Bayes |

Xu et al. [11] applied ML to develop an online framework on virtual memory access patterns for the purpose of hardware-assisted malware detection. The authors used SVM, logistic regression and random forest classifiers. Their framework performed well on the RIPE benchmark suite, reporting a 99% true positive detection rate against kernel-level rootkits and a 94% true positive detection rate against user-level memory corruption attacks. One of the paper's key insights was that, by changing control flows and data structures, malware left 'fingerprints' on program memory accesses, which led to novel ways of collecting and summarising per-function/system-call memory access patterns.

In [12], the authors proposed a malware detection method that generated a graph of operational codes (OpCode) within an executable file and then, using the 'Power Iteration' method, embedded the graph into eigenspace. This representation assisted with training SL classifiers such as KNN and SVM to classify each vector as malware or benign.

Hirano and Kobayashi [13] developed a novel machine learning-based ransomware and wiper malware detection approach using low-level memory access patterns only. The authors enhanced a 'thin and lightweight' live-forensic hypervisor by adding a function to collect low-level storage access patterns. Then, using three SL models (a Random Forest model, an SVM model and a KNN model), they achieved F1 scores of 95% in detecting ransomware and wiper malware and 93% in classifying the eight classes (four benign classes, three known ransomware and one wiper malware).

In 2019, Corum et al. [14] devised a new methodology for the detection of PDF malware using image visualisation and processing techniques. The authors chose to focus on the PDF file format due to its popularity among malware creators, as it provides flexibility in embedding various types of content. After converting PDF files to grayscale images and extracting relevant features (keypoint descriptors and texture features), the authors applied three SL algorithms (Random Forests, Decision Trees and K-Nearest Neighbour) to create a classification model of each PDF file (represented as a feature vector). This resulted in F1 scores of up to 99.48%, comparable to popular antivirus scanners such as Microsoft, Symantec and nProtect. Additionally, the method proved more robust at resisting reverse mimicry attacks than PDF Slayer, a top learning-based method for PDF malware detection.

A more recent paper [15] introduced a malware detection model based on API call sequence analysis using a KNN algorithm for malware classification and detection. The model demonstrated an accuracy of 98.17% and was also found to be effective at real-time intrusion detection on PCs.

Another recent approach [16] used Gradient-Boosting Decision Trees with a customised log loss function on the FFRI Dataset 2019, showing an 82% reduction in the number of false positives.

Table II outlines the most recent and noteworthy studies using SL-based techniques, presenting the main characteristics of malware detection methodologies, as well as detailing the AI models used and features extracted by each model.

Table II

Characteristics of Shallow Learning-Based Methods for Malware Detection

|  |  | 2017 | | 2019 | 2021 | | 2022 |
| --- | --- | --- | --- | --- | --- | --- | --- |
|  |  | [11] | [12] | [14] | [15] | [16] | [13] |
| SL Models | KNN |  | X | X | X |  | X |
|  | LR | X |  |  |  |  |  |
|  | Random Forest | X |  | X |  |  | X |
|  | Decision Trees |  |  | X |  | X |  |
|  | SVM | X | X |  |  |  | X |
| Features | API Calls |  |  |  | X |  | X |
|  | OpCode graph |  | X |  |  |  |  |
|  | Keypoint Descriptors |  |  | X |  |  |  |
|  | Texture Features |  |  | X |  |  |  |
|  | Memory Features | X |  |  |  |  | X |
|  | PE Header |  |  |  |  | X |  |

## 6. Deep Learning Classification Methods for Malware Detection

Deep learning (DL) includes machine learning methods based on artificial neural networks (algorithms designed to work like the human brain) with representation learning, which enable the network to learn from unsupervised data and solve complex problems [17]. Deep learning has been used in cyber security for a variety of applications, successfully improving accuracy, performance, reliability and scalability when applied in real-time [17]. Notably, the majority of recent malware detection techniques use DL, as will become apparent from the data presented in Table IV.

Deep learning techniques were introduced in malware detection to address the shortcomings of using SL classifiers – i.e., the dependency on manually selecting features for training and testing of the model, time constraints (SL classifiers are more time-consuming) and the limitations in processing large datasets.

Fig. 2 below lists the most commonly used DL classification methods.

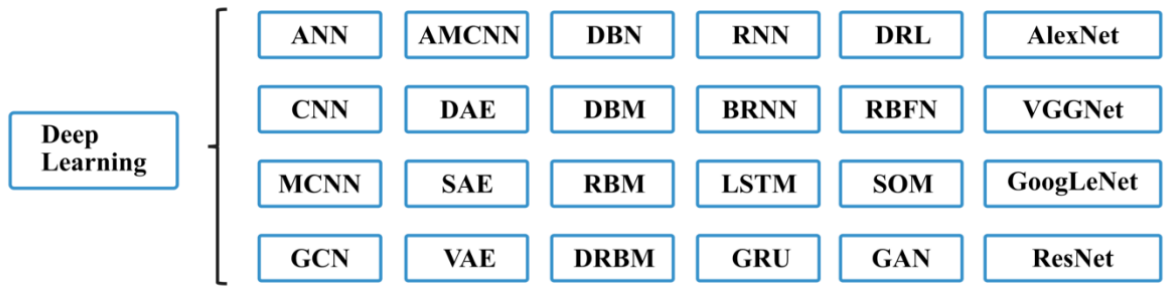

Fig. 2. Common deep learning methods

Table III outlines all the acronyms referenced in Fig. 2.

Table III
Acronyms used in Fig. 2

| ANN | Artificial Neural Network | RNN | Recurrent Neural Network |
|---|---|---|---|
| CNN | Convolutional Neural Network | BRNN | Bidirectional Recurrent Neural Network |
| MCNN | Multichannel Convolutional NN | LSTM | Long Short-Term Memory |
| GCN | Graph Convolutional Networks | GRU | Gated Recurrent Units |
| AMCNN | Attention-based Multichannel Convolutional Neural Network | DRL | Deep Reinforcement Learning |
| DAE | Denoising Autoencoders | RBFN | Radial Basis Function Networks |
| SAE | Sparse Autoencoder | SOM | Self-Organising Map |
| VAE | Variational Autoencoders | GAN | Generative Adversarial Network |
| DBN | Deep Belief Network | AlexNet | CNN designed by Alex Krizhevsky |
| DBM | Deep Boltzmann Machine | VGGNet | Visual Geometry Group NN |
| RBM | Restricted Boltzmann Machine | GoogLeNet | 22-layer CNN designed by Google researchers |
| DRBM | Discriminative Restricted Boltzmann Machine | ResNet | Residual Neural Network |

A paper published in 2018 [18] proposed a heterogeneous deep learning framework called 'DeepAM' for malware detection. This framework comprised an AutoEncoder stacked with multilayered RMBs and an additional layer of associative memory, designed to perform unsupervised feature learning and supervised fine-tuning for the detection of unknown malware.

Abdelsalam et al. [19] used convolutional neural networks for malware detection in cloud infrastructures. The approach initially involved using a 2D CNN to train on metadata obtained from a hypervisor – from Virtual Machines running various malware (Trojans and rootkits, primarily). Subsequently, a 3D CNN was used to enhance the classifier accuracy of the 2D CNN and to significantly reduce mislabelled samples during data collection and training, resulting in a 79% accuracy for the 2D CNN model and overall accuracy of 90% reached by adding the 3D CNN model.

In [20], the authors presented a Long Short-Term Memory (LSTM) network-based algorithm aiming to bypass the multiclass imbalance problem inherent in DGA malware detection. DGA is an abbreviation for 'Domain Generation Algorithm' – a piece of malicious code designed to create domain names that can be used as a command-and-control (C&C) communication channel with the cybercriminal launching the attack. The improved LSTM algorithm was, therefore, built to combine both binary and multiclass classification models and adapted to be cost-sensitive – an essential requirement in real-world detection scenarios. The authors demonstrated a 7% increase in macro-averaging recall and precision compared to the original LSTM and a number of other advanced cost-sensitive methods of DGA malware detection.

Similarly, HaddadPajouh et al. [21] utilised LSTM for malware detection but with a different aim: detecting malware in IoT devices. The Internet of Things (IoT) comprises physical objects equipped with sensors, software, processing power and other technologies that allow them to connect to other systems via the Internet or other communication means. Since IoT devices are increasingly being used in various industries and daily life, they have become a valuable target for attackers, with various emerging malware specifically designed to compromise IoT devices.

In their paper [21], the authors employed recurrent neural networks (RNNs) to examine the execution operation codes (OpCodes) of ARM-based IoT applications, then used three LSTM configurations to evaluate the trained model. The authors compared the LSTM configurations with various SL classifiers (decision trees and SVMs) and found that LSTM maintained the highest accuracy (98.18%) in malware detection.

A recent paper published in 2022 [22] proposed a new malware detection approach largely based on graph convolutional networks (GCN). This approach achieved an accuracy of 98.32%, higher than many other existing methods. The proposed system comprised four steps:

- Extracting API call sequences from the malicious code.

- Generating a directed cycle graph.
- Designing a weight model based on Markov chains and PCA to extract the graph feature map.
- Using a GCN classifier for malware detection.

The authors showed that the increased accuracy was due to the fact that, as the malware data set increased, the extracted features produced by this approach were increasingly more accurate, providing robust malware classification and detection capabilities.

Another group of researchers [23] developed a novel method of malware detection based on binary visualisation and Self-Organising Incremental Neural Networks (SOINN). The authors investigated the performance of the suggested method to discover harmful payloads in various file types. Their experimental findings revealed 91.7% and 94.1% detection accuracies for ransomware in .pdf and .doc files, respectively.

Maniriho et al. [24] focused on known and zero-day malware attacks on Windows devices, which have arguably become the primary target of malware attacks in recent years. Using a dynamic analysis approach, the authors introduced a novel behavioural dataset called MalBehavD-V1, consisting of Windows API calls extracted from benign and malicious executable files.

Furthermore, the authors developed MalDetConv, a behaviour-based malware detection system employing a text processing-based encoder to convert API call features into a format that deep learning models could use. MalDetConv then used an automatic feature extractor created as a hybrid of a convolutional neural network and bidirectional gated recurrent unit (CNN-BiGRU) to select high-level features of the API Calls. These features were then fed into a fully connected neural network module for malware classification. Additionally, in order to facilitate security analysts' task of evaluating malware threats, MalDetConv also used an explainable component to enumerate the features that contributed to the classification outcome. The proposed system achieved accuracies of 95.73%, 96.10%, 98.18%, and 99.93% on the detection of previously unknown malware from Allan and John, MalBehavD-V1, Brazilian, and Ki-D datasets, respectively.

Many of the deep learning models described thus far are large and complex in scale and structure, making the task of retraining them time and computation-intensive. Retraining models is frequently necessary due to the capability of malware to rapidly evolve to avoid detection, rendering older models obsolete with regard to detection accuracy. Additionally, as almost any electronic device is vulnerable to malware and many such devices lack the computational capacity to run complex detection systems, an emphasis on lightweight solutions is required.

To address this need, a team of researchers created a deep Convolutional Neural Network (CNN) named SeqNet [25], featuring a considerably smaller-sized detection model, a shorter training process and an enhanced potential for retraining. The authors demonstrated that SeqNet maintained detection accuracy even when substantially reducing the number of parameters – by avoiding contextual confusion and reducing semantic loss.

Table IV presents the latest and most significant papers describing DL techniques for malware detection.

Table IV

Characteristics of Deep Learning Approaches for Malware Detection

|  |  | 2018 | | | | 2019 | 2022 | | |
| --- | --- | --- | --- | --- | --- | --- | --- | --- | --- |
|  |  | [18] | [19] | [20] | [21] | [23] | [22] | [24] | [25] |
| DL Models | CNN |  |  | X |  |  |  | X | X |
|  | GCN |  |  |  |  |  | X |  |  |
|  | AE | X |  |  |  |  |  |  |  |
|  | RNN |  |  |  | X |  |  |  |  |
|  | SOINN |  |  |  |  | X |  |  |  |
|  | RBM | X |  |  |  |  |  |  |  |
|  | GRU |  |  |  |  |  |  | X |  |
|  | LSTM |  |  | X | X |  |  |  |  |
| SL Classifiers |  |  |  |  |  | X | X |  |  |
| Features | API Calls | X |  |  |  |  | X | X |  |
|  | OpCode sequence |  |  |  | X |  |  |  |  |
|  | Performance Metrics |  | X |  |  |  |  |  |  |
|  | Domain name features |  |  | X |  |  |  |  |  |
|  | Multi features |  |  |  |  |  | X |  |  |
|  | No feature engineering |  |  |  |  |  |  |  | X |

## 7. Bio-Inspired Computing Methods for Malware Detection

Bio-inspired computing is a branch of AI created to mimic biological systems – in particular, the behaviour and characteristics of biological organisms – to solve a variety of computational problems. Bio-inspired computing has emerged, in recent years, as a very popular AI method, with techniques such as genetic algorithms (GAs), particle swarm optimisation (PSO), evolution strategies (ES), ant colony optimisation (ACO) and artificial immune systems (AIS) among the most common in cyber security applications [26].

In malware detection, bio-inspired algorithms have historically been used more heavily in feature selection than classification and detection.

For instance, Abbasi et al. [27] proposed an automated, wrapper-based feature selection method employing PSO for behaviour-based ransomware detection and classification.

In another example [28], the authors detailed the use of evolutionary Genetic Algorithms for discriminatory feature selection. This method resulted in a feature subset that was later used to train machine learning classifiers for Android Malware Detection, producing better accuracy (94%) with a minimal number of features, thereby reducing computational requirements.

Another group of researchers [29] used rough sets combined with an improvised PSO algorithm (called PSORS-FS) for permission-based malware detection in Android systems.

One of the most compelling developments in recent years is the field of AIS. In order to understand the history of AIS, it is essential to learn how biological immune systems (BIS) work.

A BIS is a complex system of biological components that protect an organism against infection and disease by systematically identifying and destroying pathogens – such as viruses, bacteria and parasites – as well as tumour cells. To adequately recognise such 'intruders' as distinct from the body's healthy cells, the immune system must be capable of adapting to pathogens that rapidly evolve and produce adaptations aimed at avoiding immune defences. In short, some of the basic characteristics of a BIS include self-tolerance, specificity, diversity and adaptability. These characteristics are inherited and utilised by artificial immune systems to overcome many of the challenges in cyber security [30].

Artificial Immune Systems are rule-based machine learning computational systems inspired by the field of theoretical immunology and designed to abstract the structure and function of a BIS. Popular AIS algorithms include the Danger Theory Algorithm, the Negative Selection Algorithm, the Clonal Selection Algorithm (CSA) and the Dendritic Cells Algorithm (DCA) [30].

In 2017, Brown et al. [31] proposed a Multiple-Detector Set Artificial Immune System (mAIS), which improved upon existing AIS systems by evolving multiple-detector sets concurrently via negative selection. This approach achieved a 93.33% accuracy on a dataset of information flows captured from malicious and benign Android applications, outperforming eight other AIS and mAIS methods.

The authors in [32] developed an innovative 'Artificial Evolutionary Fuzzy LSTM Immune System' trained on raw bytes of an executable file by using a heuristic-based machine learning method combining Long-Short-Term Memory (LSTM), evolutionary intelligence and fuzzy knowledge to abstract the function of a BIS. This novel method achieved an accuracy of 98.59% in detecting Portable Executable Malware.

In [33], a Negative-Positive-Selection (NPS) method was proposed, using an AIS for malware detection in IoT devices. The performance of this new NPS method was benchmarked against other IoT malware detection methodologies and demonstrated a 21% improvement in malware detection and a 65% reduction in the number of detectors necessary, with overall reduced computational requirements – making it highly suitable for IoT devices.

In Table V, the most recent research papers covering bio-inspired computing in malware detection are outlined.

Table V
Characteristics of Bio-Inspired Computing Methods for Malware Detection

|  |  | 2017 | 2019 | 2019 | 2021 | 2021 | 2022 |
|---|---|---|---|---|---|---|---|
|  |  | [31] | [28] | [29] | [32] | [33] | [27] |
| Bio-Inspired Models | GA |  | X |  |  |  |  |
|  | ES |  |  |  | X |  |  |
|  | PSO |  |  | X |  |  | X |
|  | AIS |  |  |  |  | X |  |
|  | NPS |  |  |  |  | X |  |
|  | mAIS | X |  |  |  |  |  |
| ML Classifiers |  |  | X |  | X |  |  |

## 8. Android Malware Detection

The reliance on mobile phones is now greater than ever, becoming an inherent part of our everyday lives. In 2021, a total of 1.43 billion smartphones were manufactured [34], and Android dominated 71.5% of the market share [35]. Due to the increasing development of intelligent mobile terminals, Android has become the most popular smartphone platform; however, its open-source nature is leading to increased vulnerabilities exploitable by malware attacks.

In 2018, Li et al. [36] introduced a new Android malware detection system called SIGPID (Significant Permission Identification), aiming to achieve the highest possible accuracy while analysing the minimum number of permissions. Rather than extracting and analysing all Android permissions, SIGPID used three levels of pruning (permission ranking with negative rate, permission mining with association rules and support-based permission ranking) to mine permission data in order to identify only those permissions that were relevant in detecting malicious apps. SIGPID then used various machine learning classification approaches, of which SVM was the most accurate, reaching over 90% in precision, recall, accuracy, and F-measure. Overall, SIGPID proved more effective than other state-of-the-art malware detection techniques (such as DREBIN and Permission-Induced Risk Malware Detection), detecting 93.62% of malicious apps in the data set and 91.4% of unknown malware. The authors also tested the SIGPID approach with 67 commonly used supervised learning algorithms, showing that 55 out of 67 algorithms were capable of achieving F-measures of at least 85%.

In another method for Android OS malware detection [37], the authors built deep learning Gated Recurrent Unit (GRU) architectures of Recurrent Neural Networks (RNN). Their model, along with other traditional deep learning models, was trained on the CICAndMal2017 dataset, consisting of realistic samples of malware and benign Android applications. The proposed deep learning classifier reached 98.2% accuracy, outperforming the other traditional classifiers used in Android malware detection.

Saif et al. [38] developed a framework based on Deep Belief Networks for Android-based malware detection systems, combining high-level static analysis, dynamic analysis and system calls in feature extraction. The framework achieved an accuracy of 99.1%, outperforming some of the classical ML techniques.

The DL model proposed by the authors in [39] examined how a CNN-on-Matrix approach could be used to detect malware in Android OS. Named AdMat, their model interpreted and treated apps as images. The images were then stored in matrices at a compressed resolution of 219 x 219 to increase data processing efficiency. The resulting matrices were then used as inputs for the CNN, and the model was trained to classify malware and benign apps. AdMat demonstrated an average detection rate of 98.26%, with a 97% accuracy in classifying malware families. However, some limitations were highlighted, such as a lack of robustness against dynamic code loading attacks due to its use of static analysis only and a trade-off between the mode's performance and the number of features used.

The authors in [40] presented an Android malware detection system called MaMaDroid, which abstracted API calls performed by an app to a class, family or package. The resulting API call graph was modelled as Markov chains, improving the model's resiliency to API changes. Classification was performed using four SL algorithms: 1-Nearest Neighbour (1-NN), 3-Nearest Neighbour (3-NN), RF and SVM, with RF demonstrating the highest accuracy. The authors also showed that MaMaDroid effectively detected unknown malware samples with a date of creation earlier or similar to the samples on which it was trained, reaching an F-measure as high as 99%.

Another group [41] explored the effectiveness of using features selected from Android manifest file permissions (analysed with the help of APKTool for decompilation support), together with four SL classifiers (Random Forest, Support Vector Machine, Gaussian Naïve Bayes, and K-Means) used to classify Android apps as malicious or benign. Of the four algorithms, Random Forest achieved the best performance, with 82.5% precision and 81.5% accuracy. The authors also compared their approach against readily available antivirus engines, showing significant improvements with all four SL classifiers.

Table VI provides an overview of the latest papers on Android malware detection, highlighting SL/DL classifiers used, as well as relevant features extracted/selected.

Table VI

Characteristics of Android Malware Detection approaches

| | | 2016 | 2018 | | 2021 | | |
|---|---|---|---|---|---|---|---|
| | | [40] | [36] | [38] | [37] | [39] | [41] |
| SL Classifiers | 1-NN | X | | | | | |
| | 3-NN | X | | | | | |
| | RF | X | | | | | X |
| | SVM | X | X | | | | X |
| | GNB | | | | | | X |
| | K-Means | | | | | | X |
| DL Classifiers | CNN | | | | | X | |
| | DBN | | | X | | | |
| | RNN | | | | X | | |
| | GRU | | | | X | | |
| Features | API Calls | X | | | X | X | |
| | Permissions | | X | | X | | X |
| | System calls | | | X | | | |

## 9. AI-Assisted Malware

As defences against malware attacks become more sophisticated, intelligent and efficient, cybercriminals have continuously developed increasingly creative means of evasion. One of the main priorities of modern malware is to remain invisible to detection and sidestep anti-malware solutions. Given the rapid progress of AI, it is inevitable that cybercriminals will implement this technology in evasion techniques. Furthermore, malware developers could adopt evolution-based algorithms to create novel malware strains with the ability to adapt easily to new environments.

With AI being integrated into defences against malware attacks, cybercriminals may employ any means necessary to disrupt these efforts. Techniques such as poisoning datasets used by AI [42, 43] or reverse engineering ML models could be used to surpass security solutions.

Rigaki and Garcia [44] adopted DL techniques such as Generative Adversarial Networks (GANs) to create malicious malware samples that managed to avoid detection by simulating the behaviours of legitimate applications. Another approach by researchers in [45] suggested a DL technique in which malware was capable of identifying its target using voice recognition, geological location and facial recognition before attacking.

As a further improvement, the authors in [46] fused a neural network and swarm-based intelligence with a common computer virus to create a neural swarm virus with enhanced evasion capability.

Combined, these trends suggest an alarming tendency towards using AI techniques for malicious purposes.

Therefore, future research in the field of malware detection should rely heavily on such technologies in order to hinder cybercriminals' progress and achieve detection accuracies which are orders of magnitude greater than those in the past with regard to both recognising unknown malware and doing so at increasingly lower computational demands and lower costs.

## 10. Conclusion

As technology continues to advance, new challenges for malware detection and security will undoubtedly emerge. The increasing computational complexity of cyber-attacks creates a need for revolutionary solutions – ones that are more flexible, effective and robust. This literature review focused on the state-of-the-art implementations of AI-based techniques for malware detection in PC, cloud, Android and IoT environments, with a deep dive into the intelligent algorithms used in each analysed research paper. The AI methods outlined in this review centred around shallow learning-based classification algorithms, deep learning and bio-inspired computing. A background of the use of AI in creating ever more advanced malware was also provided, highlighting how cybercriminals are rushing to both disrupt defences and enhance malware but also providing a reminder that by researching novel ways to detect evolving malware, such malicious efforts can be counteracted.

**References**

[1] M. Fichtenkamm, G. Burch, J. Burch, *"Cybersecurity in a COVID-19 World: Insights on How Decisions Are Made."* libraryguides.vu.edu.au. https://www.isaca.org/resources/isaca-journal/issues/2022/volume-2/cybersecurity-in-a-covid-19-world (accessed September 12, 2022).

[2] S. Morgan, *"Cybercrime to Cost the World $10.5 Trillion Annually by 2025."* cybersecurityventures.com https://cybersecurityventures.com/hackerpocalypse-cybercrime-report-2016/ (accessed September 12, 2022).

[3] Deloitte, *"Cyber crime – the risks of working from home."* deloitte.com https://www2.deloitte.com/ch/en/pages/risk/articles/covid-19-cyber-crime-working-from-home.html (accessed September 13, 2022).

[4] C. Nabe, *"Impact of COVID-19 on Cybersecurity."* deloitte.com https://www2.deloitte.com/ch/en/pages/risk/articles/impact-covid-cybersecurity.html (accessed September 14, 2022).

[5] Markets and Markets, *"AI in Cybersecurity Market by Offering (Hardware, Software, Service), Technology (Machine Learning, Context Awareness, NLP), Deployment Type, Security Type, Security Solution, End-user, and Geography - Global Forecast to 2025."* marketsandmarkets.com https://www.marketsandmarkets.com/market-reports/ai-in-cybersecurity-market-224437074.html (accessed September 19, 2022).

[6] G. Belani, "The Use of Artificial Intelligence in Cybersecurity: A Review." computer.org https://www.computer.org/publications/tech-news/trends/the-use-of-artificial-intelligence-in-cybersecurity (accessed September 20, 2022).

[7] A. Ray, *Cybersecurity for Connected Medical Devices,* United States of America: Academic Press, 2021, pp. 217-262.

[8] Forrester, *"The Forrester Wave: Endpoint Detection and Response Providers, Q2 2022."* forrester.com https://www.forrester.com/report/the-forrester-wave-tm-endpoint-detection-and-response-providers-q2-2022/RES176332?reference=twitter&utm_source=twitter&utm_medium=ppc&utm_campaign=msbg_cx_cert (accessed September 20, 2022).

[9] F. A. Aboaoja, A. Zainal, F. A. Ghaleb, B. A. S. Al-rimy, T. A. E. Eisa, and A. A. H. Elnour, "Malware Detection Issues, Challenges, and Future Directions: A Survey," *Applied Sciences*, vol. 12, no. 17, p. 8482, 2022.

[10] Y. Xu, Y. Zhou, P. Sekula, and L. Ding, "Machine learning in construction: From shallow to deep learning," *Developments in the built environment*, vol. 6, p. 100045, 2021.

[11] Z. Xu, S. Ray, P. Subramanyan, and S. Malik, "Malware detection using machine learning based analysis of virtual memory access patterns," in *Design, Automation & Test in Europe Conference & Exhibition (DATE), 2017*, 2017, pp. 169–174.

[12] H. Hashemi, A. Azmoodeh, A. Hamzeh, and S. Hashemi, "Graph embedding as a new approach for unknown malware detection," *Journal of Computer Virology and Hacking Techniques*, vol. 13, no. 3, pp. 153–166, 2017.

[13] M. Hirano and R. Kobayashi, "Machine Learning-based Ransomware Detection Using Low-level Memory Access Patterns Obtained From Live-forensic Hypervisor," *arXiv preprint arXiv:2205.13765*, 2022.

[14] A. Corum, D. Jenkins, and J. Zheng, "Robust PDF malware detection with image visualization and processing techniques," in *2019 2nd International Conference on Data Intelligence and Security (ICDIS)*, 2019, pp. 108–114.

[15] T. A. Assegie, "An optimized KNN model for signature-based malware detection," *Tsehay Admassu Assegie." An Optimized KNN Model for Signature-Based Malware Detection". International Journal of Computer Engineering In Research Trends (IJCERT), ISSN*, pp. 2349–7084, 2021.

[16] Y. Gao, H. Hasegawa, Y. Yamaguchi and H. Shimada, "Malware Detection Using Gradient Boosting Decision Trees with Customized Log Loss Function," 2021 International Conference on Information Networking (ICOIN), 2021, pp. 273-278, doi: 10.1109/ICOIN50884.2021.9333999.

[17] P. Dixit and S. Silakari, "Deep learning algorithms for cybersecurity applications: A technological and status review," *Computer Science Review*, vol. 39, p. 100317, 2021.

[18] Y. Ye, L. Chen, S. Hou, W. Hardy, and X. Li, "DeepAM: a heterogeneous deep learning framework for intelligent malware detection," *Knowledge and Information Systems*, vol. 54, no. 2, pp. 265–285, 2018.

[19] M. Abdelsalam, R. Krishnan, Y. Huang, and R. Sandhu, "Malware detection in cloud infrastructures using convolutional neural networks," in *2018 IEEE 11th International conference on cloud computing (CLOUD)*, 2018, pp. 162–169.

[20] D. Tran, H. Mac, V. Tong, H. A. Tran, and L. G. Nguyen, "A LSTM based framework for handling multiclass imbalance in DGA botnet detection," *Neurocomputing*, vol. 275, pp. 2401–2413, 2018.

[21] H. HaddadPajouh, A. Dehghantanha, R. Khayami, and K.-K. R. Choo, "A deep recurrent neural network based approach for internet of things malware threat hunting," *Future Generation Computer Systems*, vol. 85, pp. 88–96, 2018.

[22] S. Li, Q. Zhou, R. Zhou, and Q. Lv, "Intelligent malware detection based on graph convolutional network," *The Journal of Supercomputing*, vol. 78, no. 3, pp. 4182–4198, 2022.

[23] I. Baptista, S. Shiaeles, and N. Kolokotronis, "A novel malware detection system based on machine learning and binary visualization," in *2019 IEEE International Conference on Communications Workshops (ICC Workshops)*, 2019, pp. 1–6.

[24] P. Maniriho, A. N. Mahmood, and M. J. M. Chowdhury, "MalDetConv: Automated Behaviour-based Malware Detection Framework Based on Natural Language Processing and Deep Learning Techniques," *arXiv preprint arXiv:2209.03547*, 2022.

[25] J. Xu, W. Fu, H. Bu, Z. Wang, and L. Ying, "SeqNet: An Efficient Neural Network for Automatic Malware Detection," *arXiv preprint arXiv:2205.03850*, 2022.

[26] S. N. Mthunzi, E. Benkhelifa, T. Bosakowski, and S. Hariri, "A bio-inspired approach to cyber security," in *Machine Learning for Computer and Cyber Security*, CRC Press, 2019, pp. 75–104.

[27] M. S. Abbasi, H. Al-Sahaf, M. Mansoori, and I. Welch, "Behavior-based ransomware classification: A particle swarm optimization wrapper-based approach for feature selection," *Applied Soft Computing*, vol. 121, p. 108744, 2022.

[28] A. Fatima, R. Maurya, M. K. Dutta, R. Burget, and J. Masek, "Android malware detection using genetic algorithm based optimized feature selection and machine learning," in *2019 42nd International conference on telecommunications and signal processing (TSP)*, 2019, pp. 220–223.

[29] A. Bhattacharya, R. T. Goswami, and K. Mukherjee, "A feature selection technique based on rough set and improvised PSO algorithm (PSORS-FS) for permission based detection of Android malwares," *International journal of machine learning and cybernetics*, vol. 10, no. 7, pp. 1893–1907, 2019.

[30] M. A. M. Ali and M. A. Maarof, "Malware detection techniques using artificial immune system," in *Proceedings of the International Conference on IT Convergence and Security 2011*, 2012, pp. 575–587.

[31] J. Brown, M. Anwar, and G. Dozier, "An artificial immunity approach to malware detection in a mobile platform," *EURASIP Journal on Information Security*, vol. 2017, no. 1, pp. 1–10, 2017.

[32] J. Jiang and F. Zhang, "Detecting Portable Executable Malware by Binary Code Using an Artificial Evolutionary Fuzzy LSTM Immune System," *Security and Communication Networks*, vol. 2021, 2021.

[33] H. Alrubayyi, G. Goteng, M. Jaber, and J. Kelly, "A novel negative and positive selection algorithm to detect unknown malware in the IoT," in *IEEE INFOCOM 2021-IEEE Conference on Computer Communications Workshops (INFOCOM WKSHPS)*, 2021, pp. 1–6.

[34] Statista, *"Number of smartphones sold to end users worldwide from 2007 to 2021"* statista.com https://www.statista.com/statistics/263437/global-smartphone-sales-to-end-users-since-2007/ (accessed October 5, 2022).

[35] StatCounter, *"Mobile Operating System Market Share Worldwide"* gs.statcounter.com https://gs.statcounter.com/os-market-share/mobile/worldwide (accessed October 5, 2022).

[36] J. Li, L. Sun, Q. Yan, Z. Li, W. Srisa-An, and H. Ye, "Significant permission identification for machine-learning-based android malware detection," *IEEE Transactions on Industrial Informatics*, vol. 14, no. 7, pp. 3216–3225, 2018.

[37] O. Elayan and A. Mustafa, 'Android Malware Detection Using Deep Learning', Procedia Computer Science, vol. 184, pp. 847–852, 01 2021.

[38] D. Saif, S. El-Gokhy, and E. Sallam, "Deep Belief Networks-based framework for malware detection in Android systems," *Alexandria engineering journal*, vol. 57, no. 4, pp. 4049–4057, 2018.

[39] L. N. Vu and S. Jung, "AdMat: A CNN-on-matrix approach to Android malware detection and classification," *IEEE Access*, vol. 9, pp. 39680–39694, 2021.

[40] E. Mariconti, L. Onwuzurike, P. Andriotis, E. De Cristofaro, G. Ross, and G. Stringhini, "Mamadroid: Detecting android malware by building markov chains of behavioral models," *arXiv preprint arXiv:1612.04433*, 2016.

[41] N. Herron, W. B. Glisson, J. T. McDonald, and R. K. Benton, "Machine learning-based android malware detection using manifest permissions," 2021.

[42] P. Li, Q. Liu, W. Zhao, D. Wang, and S. Wang, "Bebp: an poisoning method against machine learning based idss," *arXiv preprint arXiv:1803.03965*, 2018.

[43] S. Chen *et al.*, "Automated poisoning attacks and defenses in malware detection systems: An adversarial machine learning approach," *computers & security*, vol. 73, pp. 326–344, 2018.

[44] M. Rigaki and S. Garcia, "Bringing a gan to a knife-fight: Adapting malware communication to avoid detection," in *2018 IEEE Security and Privacy Workshops (SPW)*, 2018, pp. 70–75.

[45] M. P. Stoecklin, "Deeplocker: How AI can power a stealthy new breed of malware," *Security Intelligence, August*, vol. 8, 2018.

[46] T. C. Truong, J. Plucar, Q. B. Diep, and I. Zelinka, "X-ware: a proof of concept malware utilizing artificial intelligence," *International Journal of Electrical & Computer Engineering (2088-8708)*, vol. 12, no. 2, 2022.